\documentclass[article,twocolumn,amsmath,floats,showpacs,nofootinbib]{revtex4}
\usepackage{graphicx}
\usepackage{textcomp}

\usepackage{hyperref}

\hypersetup{colorlinks=true}

\begin{document}

\title{Reversible transitions in high-$T_c$ cuprates based point contacts}
\author{L.F. Rybaltchenko**, N.L. Bobrov**, V.V. Fisun**, I.K. Yanson**, A.G.M. Jansen*, and P. Wyder*}
\affiliation{High Magnetic Field Laboratory, Max-Planck-Institut f$\ddot{u}$r Festk$\ddot{o}$rperforschung and Centre National
de la Recherche Scientifique, B.P. 166, 38042 Grenoble Cedex 9, France*,\\
B. Verkin Institute for Low Temperature Physics and Engineering, 47 Lenin Avenue, 310164 Kharkov, Ukraine**\\
Email address: bobrov@ilt.kharkov.ua}
\published {\href{http://link.springer.com/article/10.1007\%2Fs100510050876}{Eur. Phys. J.} B 10, 475-480 (1999)}
\date{\today}

\begin{abstract}The influence of electric felds and currents has been investigated in the high-$T_c$ superconductors
$YBaCuO$ and $BiSrCaCuO$ using a point-contact geometry with $Ag$ as the counterelectrode, which reveal
switching transitions between states of a different resistance. The origin of this effect in point contacts
is associated with electromigration of the oxygen, driven by the electric feld as well as by the currentinduced
"electron wind". The switching effect preserves its basic features at elevated temperatures up to
room temperature and in high magnetic felds up to 10~T.
\end{abstract}
\maketitle

\section{Introduction}
Compared with conventional superconductors, the electronic
(and superconducting) properties of the cuprate
superconductors can be much more affected by the application
of a direct electric feld (see \cite{1}, and Refs. therein).
For ultrathin $YBaCuO$ layers (12-100~\AA) in a metalinsulator-
superconductor junction with a gate electrode,
it was found that the normal state resistance could change
by as much as 20\% in an applied polarization $10^8 V/cm$ \cite{2}.
The mechanism for such a large electric feld effect in
cuprate superconductors remains under discussion \cite{3,4}.
One of the two proposed explanations is associated with
the conventional Coulomb interaction of the external electric
feld with the free charges resulting in a change of the
carrier concentration in a thin film. The other explanation
involves the oxygen rearrangement driven by the applied
feld (oxygen electromigration) that changes the density
of the charge carriers (holes) in the conduction band.

A significant influence on the cuprate superconductor
properties, namely on the charge carrier concentration,
can be also obtained in point contacts (PC) with metallic
conductivity. A few years ago \cite{5}, we have observed for
the first time reversible changes of the resistance of point
contacts based on single crystal $ReBa_2Cu_3O_{7-y}$ (Re = Y
or Ho) with a normal counterelectrode. The conductivity
and superconducting parameters of the PC area could be
varied very strongly by applying a voltage over the contact
in the range of a few hundreds millivolts. Above a
positive critical voltage (usually about 300~$mV$) applied
to the superconducting electrode, the superconductivity
in the contact area could be suppressed completely, accompanied
in some cases by a transition to the semiconducting
state. The state of suppressed superconductivity
in the PC area revealed a high stability. Even an exposure
of a few hours at temperatures significantly higher than
$T_c$ did not affect noticeably this normal state. A return to
the original state (\emph{i.e.}, a restoration of the superconductivity
in the PC area) took only place by applying across
the contact the opposite voltage polarity of approximately
the same value. We suggested that the observed switching
phenomenon is associated with the oxygen migration in
the PC area in an electric field. A theoretical confirmation
of this explanation was found in papers of Chandrasekhar
et al. \cite{6}, where the authors have simulated the oxygen
dynamics in the chains of 1-2-3 type cuprate superconductors
and found a high oxygen mobility in an electric
field.

In the present work we have further investigated the
switching phenomena in point contacts with $YBaCuO$ and
$BiSrCaCuO$ samples under applied electric and magnetic
fields. We have found for the first time that in the 1-2-
3 system it is possible to suppress superconductivity in
the contact area for both electric field polarities. We assume
that apart from a direct action of the electric field
on the oxygen migration, also the current-induced "electron
wind" \cite{7} plays a role in the observed phenomena.
Because of the difference in charge between the oxygen
ions and the hole type of charge carriers, these driving
forces should act in opposite directions. The sample quality
has an influence on the efficiency of the "electron wind"
force with respect to the electric field induced migration.
For the Bi-compound of 2212 type, we observed for the
first time a moderate switching effect of one polarity only,
indicating for the presence of the direct field action on the
oxygen ions. Reduced concentration and short mean-freepath
of the holes are possible reasons of the inefficiency of
the "electron wind" in this compound.

Previous study of the magnetic field influence on the
switching effect in single crystals of 1-2-3 type with optimal
oxygen content has given ambiguous results \cite{8}. Most
of the contacts did not show any influence of the field on
the switching effect, but for some ones a strong change in
contact resistance was found. In the present work for polycrystalline
$YBaCuO$ samples with reduced oxygen content
(oxygen index 6.75), we have found that the switching effect
persists in high magnetic fields revealing only small
variations of the point contact resistance under the applied
field. A suggestion is made to connect these variations with
the inhomogeneous crystalline structure of the given samples.
At elevated temperatures (up to room temperature),
the critical switching voltage for the oxygen deficient samples
dropped by a factor of two. Such behavior supports
the oxygen electromigration model of the effects studied.
\section{Experimental features}
We have studied the switching effects in point contacts
with the high-$T_c$ superconductors $YBaCuO$ (with different
oxygen content) and $BiSrCaCuO$ (fully oxidized). The
contacts were created by gently touching a sharpened $Ag$
electrode to a fresh fracture of the superconducting electrode.
The point-contact characteristics, $I(V)$ curves and
their first derivatives $dV/dI(V)$, were measured by the
standard methods at slow sweep rates: each full voltage
span takes about a few minutes. The temperatures could
be varied over a wide range (1,5-300~$K$). In general, the
magnetic fields up to 10~T were aligned approximately
perpendicularly to the PC axis.

Typical PC resistances obtained on the samples under
study varied from several to several tens of ohms. An estimation
of the contact size \emph{d} (diameter of the conducting
constriction) according to the Maxwell formula \cite{9} gives
$d\sim 10-100~nm$, if one takes the residual resistivity value
$\rho_0\sim0.1~m\Omega~cm$
 for high quality samples of the 1-2-3
type. For the maximal voltage of 3~$V$ that could be applied
across the contact without contact destruction, the
maximal electric field applied to the contact area is about
$10^6~V/cm$, i.e. much less than the corresponding values
attained in experiments with a source-drain-gate arrangement
on a thin film with electric field effects at $10^8~V/cm$
(see, for example, \cite{2}). Even for such elevated fields in the
experiments on thin film structures, the changes of the
sample resistance did not exceed 20\%. In our experiments
the PC resistance variations could attain more than two
orders of magnitude, in spite of the relative smallness of
the electrical field applied.
\section{Switching characteristics}
\subsection{YBaCuO system}

\begin{figure}[]
\includegraphics[width=8cm,angle=0]{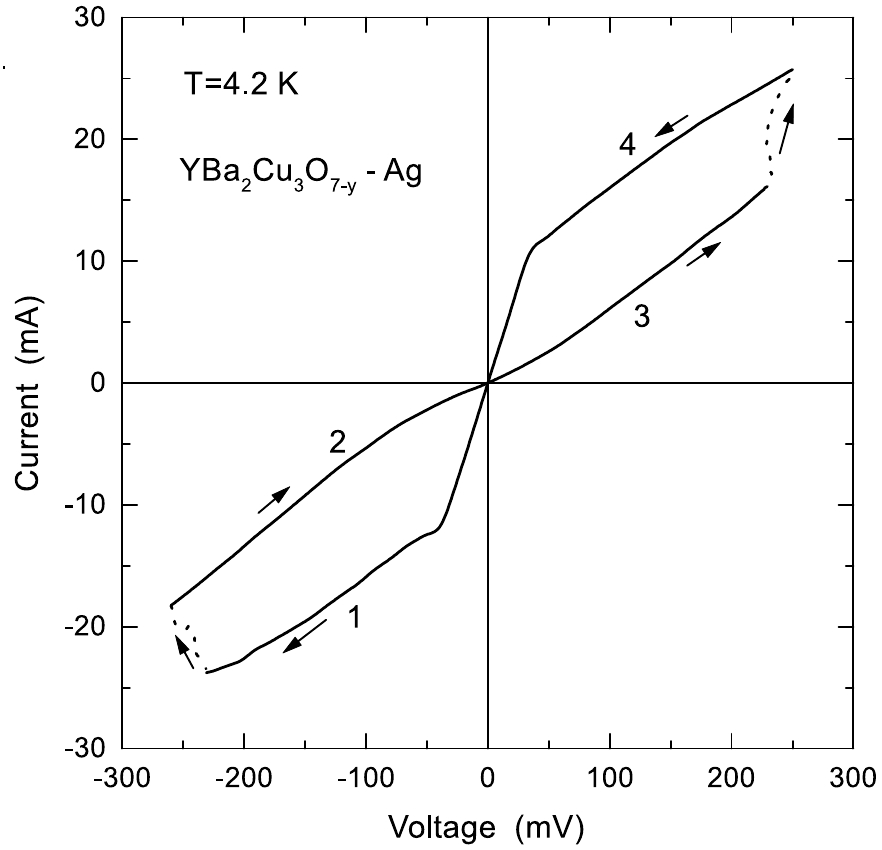}
\caption[]{Current-voltage characteristics of a $Ag-YBa_2Cu_3O_{7-y}$
point contact at 4.2~$K$ showing the direct switching effect. The
numbers and arrows near the branches denote the course of
the voltage sweep.}
\label{Fig1}
\end{figure}

Figure \ref{Fig1} represents the typical switching effect observed
in our experiments on point contacts based on single crystal
$Y$ (or $Ho$) $Ba_2Cu_3O_{7-y}$ samples.
 The arrows show the
voltage-sweep direction and the voltage sign corresponds
to the polarity of the $Ag$ electrode. As one can see, at some
negative voltage of the $Ag$ electrode, the low ohmic PC
characteristic of the superconducting state (S) (branch 1)
transforms to a high ohmic one (branch 2), corresponding
to the normal state (N). The latter, in its turn, can be
returned to the original state (transition from branch 3 to
branch 4) at the opposite electrode polarity. The shown
hysteretic $I(V)$ characteristic could be repeated over and
over again provided the amplitude of the periodic voltage
was large enough, monitoring in such way the reversible
transitions between the N-S and N-N types of point contacts.
The switching transitions between the different PC
states occurred at voltages close to a critical value $V_c$,
which for most of the single crystal samples was within
the 250-350~$mV$ range. For ceramic samples this value
could be higher by a factor 2-3, and for oxygen deficient
samples even by a factor 4-6. The $V_c$ value showed only
a weak correlation with the value of the PC differential
resistance $R_N$, measured at $eV>\Delta$ ($\Delta$ is superconducting
energy gap), which mirrors the resistance of a contact
with superconducting properties in the normal state. For
example, when $R_N$ increased by an order of the magnitude,
$V_c$ dropped only by not more then about ten percent.
We note that Figure \ref{Fig1} does not show the most extreme
PC states, which could be attained in our experiments. In
many contacts, the low voltage ($eV<\Delta$) resistance $R_0$
could change more than two orders of magnitude between
the two states.

The switching time between different PC states depends
on the voltage exceeding the critical value $V_c$. For
large voltage sweeps, for instance, of a several tens of millivolts
above $V_c$, the switching occurs faster than the time
to record the data ($\sim 0.1~s$). At voltages close to $V_c$, the
switching takes a time of about several tens of minutes \cite{5}.

The superconductivity occurred to be suppressed entirely
in the high resistance state of the contact presented
in Figure \ref{Fig1} (branches 2, 3) because the low-voltage resistance
$R_0$ is larger than the high-voltage resistance $R_N$,
\emph{i.e.}, excess currents due to Andreev reflection are absent.
As the superconductivity is suppressed for $V>V_c$ at negative
polarity of the $Ag$ electrode (left semi-plane of Fig. \ref{Fig1}),
we suggest that in given case the switching transitions are
associated with the electric field induced migration of the
negatively charged oxygen ions from the surface of the
superconducting electrode into its bulk. Furthermore, the
hysteretic behavior during the switching transitions excludes
a possibility to explain the given phenomenon by a
direct Coulomb influence on the free charge concentration.

\begin{figure}[]
\includegraphics[width=8cm,angle=0]{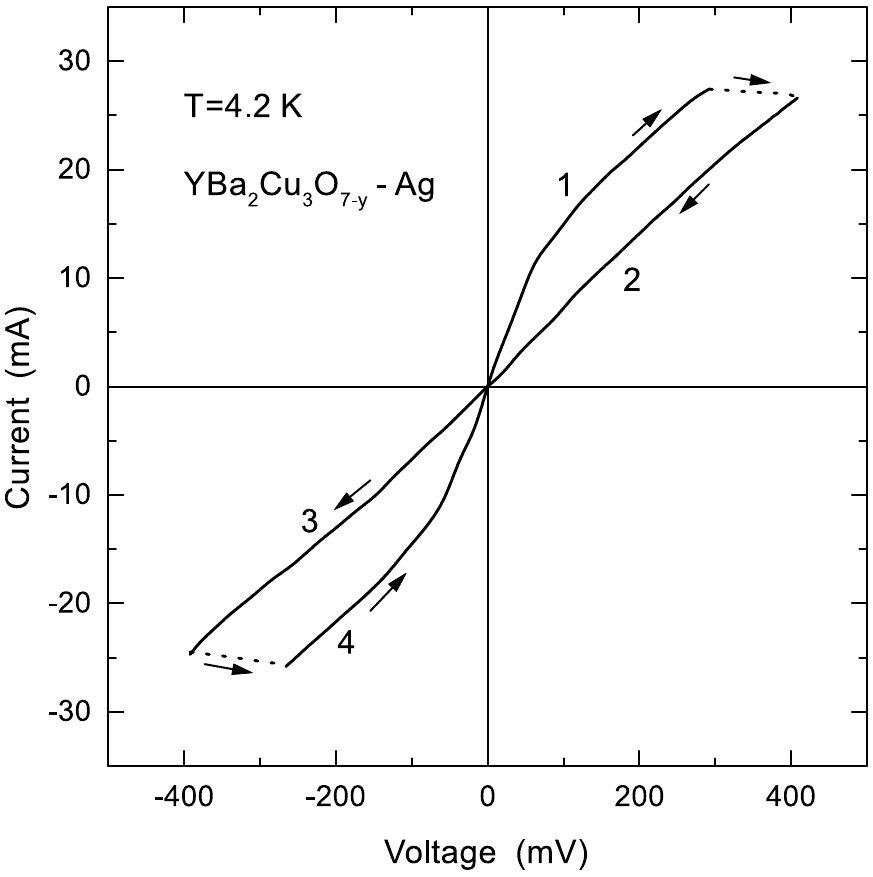}
\caption[]{Current-voltage characteristics of a point contact between
$Ag$ and single crystal $YBa_2Cu_3O_{7-y}$ at 4.2 $K$ showing
the inverse switching effect.}
\label{Fig2}
\end{figure}

In some cases, we could observe similar switching transitions
at opposite polarities (Fig. \ref{Fig2}). We call the latter
the inverse effect, in order to distinguish from the usually
observed direct effect. In our previous work \cite{5}, we
observed the inverse effect of a relatively small intensity
only together with the direct one. After a full restoration
of the superconductivity at $V>V_c$ for positive $Ag$
polarity, a subsequent voltage increase caused a partial
suppression of the superconductivity (a transition to a
higher ohmic branch of the PC $I(V)$ characteristic). Correspondingly,
during the voltage sweep in opposite direction,
the sequence of transitions was reversed, firstly the
partial suppression of the superconductivity was restored,
then the full suppression occurred. In \cite{5} we suggested
that the given phenomenon is associated with overdoping
of the PC area by the oxygen migration that could reduce
the superconducting parameters \cite{10}.

However, in the present work we have observed that in
some contacts based on single crystals the inverse effect
can be observed on its own and not on a background of the
direct effect as in previous experiments. Moreover, the inverse
effect shows a full suppression of the superconductivity
in the point-contact area (see Fig. \ref{Fig2}), and a transition
to the semiconducting state in some cases. It is reasonable
to suggest that we also deal with an oxygen migration
in the case of the inverse switching effect. The driving
force can be the current-induced "electron wind", which
due to the hole type of current carriers moves the oxygen
ions in an opposite direction compared to the electric-field
induced migration. The efficiency of the "electron-wind"
process depends on the quasiparticle momentum gain in
an applied electric field. When the mean-free-path of the
charge carriers is small, the gained momentum can be insufficient for the oxygen displacement. That is probably
the reason why the inverse effect is rarely observed in single
crystal samples, and never was observed in ceramics
and polycrystals where the mean-free-path is very short.
\subsection{BiSrCaCuO system}
In previous PC investigations of the $BiSrCaCuO$ system
\cite{11}, the switching effect has not been observed. Its observation
could be masked by the interlayer Josephson effect,
which results in many step-like structures in the $I(V)$
characteristics. On the background of such a structure, it
is difficult to recognize the features associated with the
switching effects under discussion. Nevertheless, in some
contacts based on polycrystal $(Bi_{1-x}Pb_x)_2Sr_2CaCu_2O_{8+y}$
($x\sim 0.1$) with $T_c\approx 80~K$ the step structure was suppressed,
probably due to imperfections in the layer ordering
that prevented the Josephson tunneling between $CuO_2$
layers. The partial substitution for $Bi$ by $Pb$ could be responsible
for some crystalline disorder without essential
reduction of the critical temperature.

\begin{figure}[]
\includegraphics[width=8cm,angle=0]{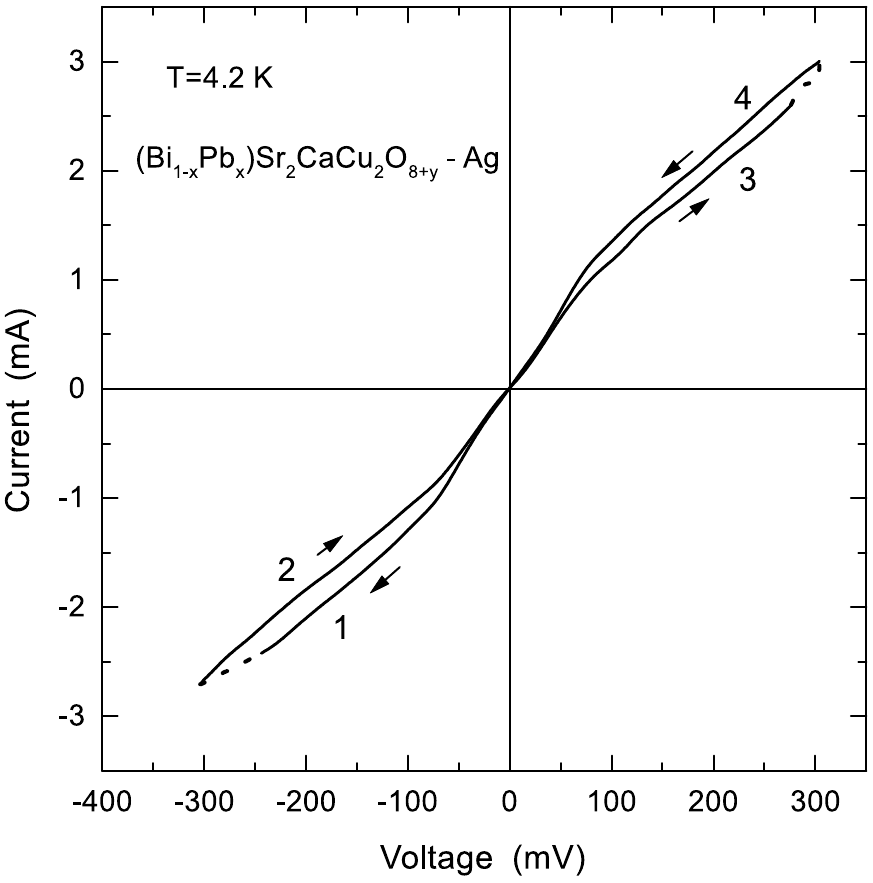}
\caption[]{Current-voltage characteristics of a point contact between
$Ag$ and polycrystal ${(Bi_{1-x}Pb_x)}_2Sr_2CaCu_2O_{8+y} (x\sim
0.1)$ at 4.2~$K$ showing the direct switching effect.}
\label{Fig3}
\end{figure}

In Figure \ref{Fig3} we present the case of a switching effect in
$(Bi_{1-x}Pb_x)_2Sr_2CaCu_2O_{8+y}$. The polarity of the switching
effect indicates that the oxygen motion in the contact area
results from the direct action of the electrostatic field, but
not from the current-induced "electron wind". The inverse
effect was not observed which can be understood as a result
of the structural disorder which limits the "electron wind"
effect. In contrast to the $Y$-cuprate system, for the
Bi-system the suppression of the superconductivity near
the contact center was relatively weak (see, for instance,
the difference between branches 1 and 2 in Fig. \ref{Fig3}). Accordingly,
the change of the resistance at zero bias $R_0$ was
also small. The small intensity of the switching effect may
be explained by the absence of the $Cu-O$ chains in the
$Bi$-cuprates that hinders the oxygen migration.

\section{Temperature influence}

\begin{figure}[]
\includegraphics[width=8cm,angle=0]{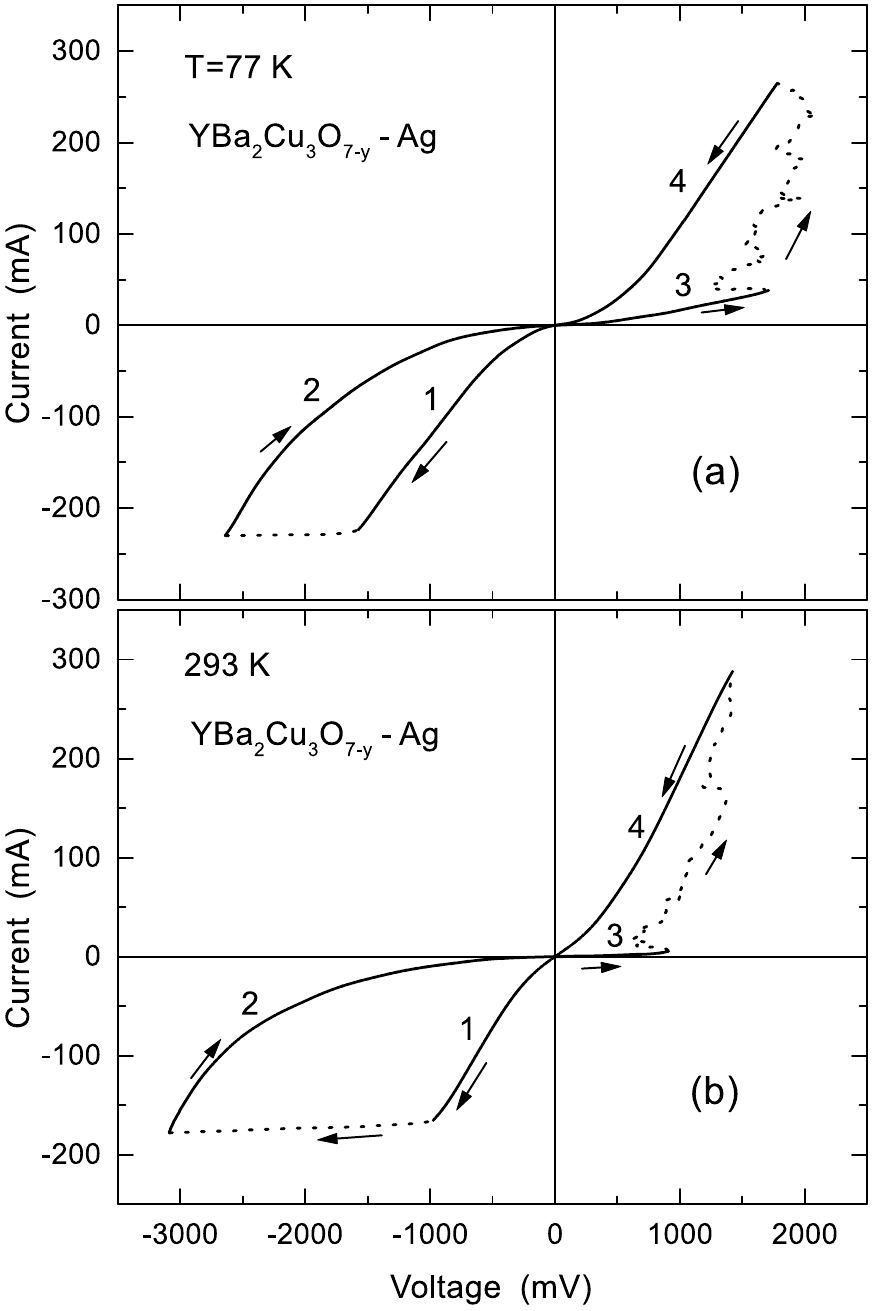}
\caption[]{Current-voltage characteristics of a point contact between
$Ag$ and oxygen deficient polycrystal $YBa_2Cu_3O_{6.75}$ at
77~$K$ (a), and at 293~$K$ (b) showing the direct switching effect.}
\label{Fig4}
\end{figure}

The observation of a strong switching effect even in oxygen
deficient $YBaCuO$ samples of low crystalline perfection
\cite{12} indicates the important role of the $Cu-O$ chains in
the oxygen electromigration in cuprate superconductors.
High intensity and stability of the switching effect in these
samples allowed the study of its temperature dependence.
In Figure \ref{Fig4} the switching characteristics in oxygen deficient $YBa_2Cu_3O_{6.75}$ are presented at 77~$K$ and at 293~$K$.
The complicated nature of the transition branch in the
right semi-plane of Figure \ref{Fig4} indicates that the transition
occurs with many elemental acts with slightly different
$V_c$, so that each of the acts is associated with the oxygen
motion in some particular structural (or substructural) element
(crystallite, block, cluster, etc.). Probably, the oxygen
deficiency defines the strong structural non-uniformity
of these samples. The transition branches at negative voltage
in Figure \ref{Fig4} do not show these features due to the use
of a current source for the recording of the $I(V)$ curves.

The absence of any observation of the inverse switching
effect in oxygen deficient samples is apparently a result of
the structural imperfection preventing a proper momentum
gain by the "electron wind". The 77~$K$ $I(V)$ characteristics
of oxygen deficient samples are not significantly
different from those recorded at 4.2~$K$, as well as the values
of a critical voltage $V_c$ of transition between different
states. However, at 293~$K$ the $V_c$ decreased essentially,
approximately by a factor 1.5-2. Besides, at 293~$K$ the
switching occurred to a higher resistance state (Fig. \ref{Fig4}b)
because at elevated temperatures the oxygen ions move
more easily in an applied electric field. These temperature
dependent data support the oxygen electromigration
model of the switching effect.

From the data presented, one can conclude that a possible
current-induced heating of the PC area does not affect
essentially the switching phenomena studied. Even
for strong temperature variations between 4.2 and 77~$K$,
the switching characteristics occurred to be similar. Earlier
\cite{12}, we have shown that these phenomena preserve
practically the same behavior in the high and low ohmic
contacts, in spite of a large difference of a electric power
dissipated in the PC area. Moreover, the switching characteristics
do not change by lowering the He bath temperature
in the superfluid phase where the heat transfer from
the PC area is significantly enhanced.
\section{Magnetic field influence}
In previous experiments devoted to a study of the magnetic
field influence on switching effects in point contacts
based on high quality single crystals $Y$ (or $Ho$)
$Ba_2Cu_3O_{7-y}$ \cite{8}, the obtained results were not unambiguous.
The current-voltage characteristics of most of the
contacts investigated remained practically unchanged in
fields up to 5~T. In some contacts, a strong change of the
PC characteristics under an applied magnetic field was
observed. A suggestion was made that such contacts contained
in the contact area twin boundaries which could be
moved in a magnetic field. According to \cite{13}, such boundaries
may be nonsuperconducting. So, their displacement
could influence the state in the PC area.

In the present investigations we studied the magnetic
field action on the switching processes in oxygen deficient
$YBa_2Cu_3O_{6.75}$ samples. In fields up to 10~T, the intensity
of the switching effect remained practically without any
essential change. The critical voltages $V_c$ of the transitions
between different contact states showed only small variations
of nonsystematic character (there were investigated
several tens of contacts).

\begin{figure}[]
\includegraphics[width=8cm,angle=0]{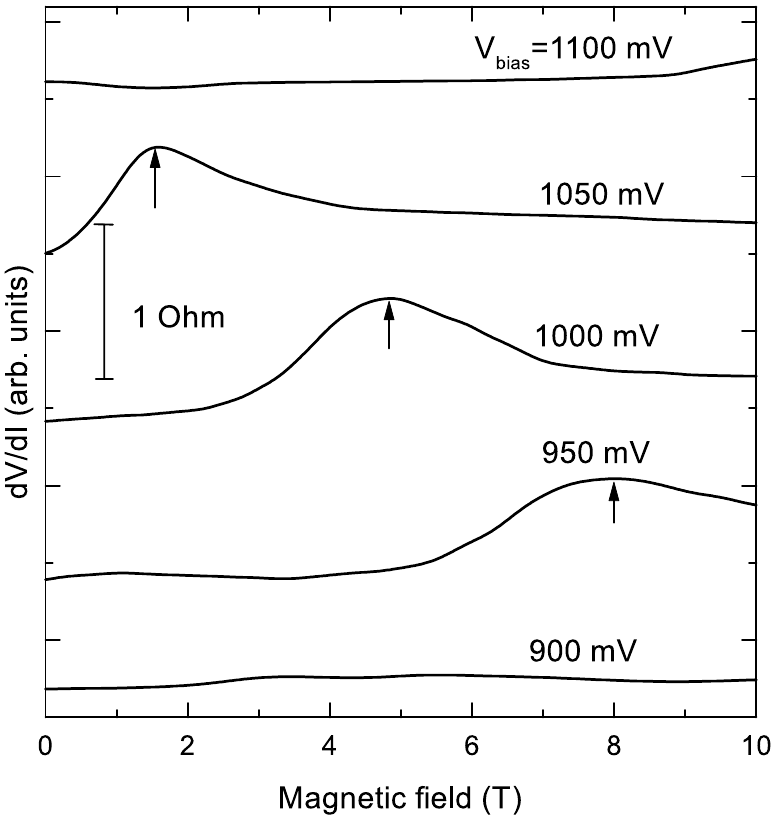}
\caption[]{The differential resistance of a point contact based on
oxygen deficient polycrystal $YBa_2Cu_3O_{6.75}\ (R(H=0)\approx 5\ \Omega
;\
$T=4.2 K) as a function of applied magnetic feld at the
indicated bias voltages. The arrows indicate the broad maxima
which shift to lower magnetic feld for higher bias voltages. For
clarity the curves are shifted vertically.}
\label{Fig5}
\end{figure}

A clear field influence was detected in the magnetic
field dependences of the differential resistance $dV/dI(H)$
recorded at fixed relatively high voltage biases $V_B$ across
the contact (Fig. \ref{Fig5}). Broad maxima usually arised in
$dV/dI(H)$ dependences in a wide voltage range, starting
from $V_B\sim 0.7~V$ and up to $V_c$ (about 1.4~$V$ at 4.2~$K$
for given samples). With $V_B$ increasing, the maximum
shifted to a smaller magnetic field range with subsequent
disappearance. Sometimes we observed two maxima simulteneously,
but mostly the succeeding maximum arised
only when the preceding one disappeared. The magnetic
field dependent contact resistance changes (10-15\%) are
much smaller than the resistance changes involved in the
switching effect. The maxima are observed for both the
high ohmic contact state and the low ohmic one.

Because given maxima arise at voltages close to $V_c$,
it is reasonable to suppose that the joint action of both
electric and magnetic fields causes a small reversible displacement
of oxygen ions over distances comparable with
the lattice parameters. Only for a subsequent voltage increase
to $V_c$, the oxygen migration would become possible
over significant distances in new metastable sites. Such
an intermediate nonsteady state of the oxygen ions may
result in a change of the contact resistance. The above
mentioned mechanism in terms of moving twin boundaries
can not probably be applied to explain the magnetic
field influence on the contact resistance in our case of the
strongly disordered oxygen deficient samples. Therefore,
we suggest another explanation.

Recently, Ratner \cite{14} has analysed electronic and magnetic
properties of the copper-oxide superconductors with
a different oxygen content. He arrived to a conclusion that
in the dielectric phase of $YBa_2Cu_3O_x$ compounds with
oxygen content close to the critical magnitude $x\sim 6.4$
(\emph{i.e.} near the dielectric-metal transition), extended ferromagnetic
clusters can be formed in the $Cu-O$ subsystem.
As one can infer from the erratic transitions between different
contact states in Figure \ref{Fig4}, the oxygen distribution
over the volume of the samples investigated is not homogeneous.
So, the occurrence of a ferromagnetic cluster in the
PC area is quite possible. In such a situation, the paramagnetic
$O^-$ ions incorporated into the cluster may be
displaced by both electric and magnetic fields, influencing
the position of neighbouring oxygen ions located in the
metallic phase and, correspondingly, the PC resistance.
\section{Summary}
In conclusion, we have found that in point contacts based
on $YBaCuO$ samples with normal counterelectrode, it is
possible to observe the switching from a superconducting
to a normal branch of the current-voltage characteristics
for both polarities of the applied voltage. The data obtained
support the validity of the oxygen electromigration
model for the explanation of electric field effects in the
point-contact geometry. A migration of the oxygen ions in
the point-contact area of the superconducting electrode
may be caused by both the direct field action and the
current-induced "electron wind" which act in opposite directions.
Depending on the local sample perfection, the
predominance of one of these two mechanisms determines
the observed polarity of the effect. The switching effect
observed in $BiSrCaCuO$ point contacts has a relatively
small intensity compared to the experiments on $YBaCuO$
which could be associated to the absence of $Cu-O$ chains
with mobile oxygen ions in this compound. The switching
field effect keeps its basic features in high magnetic fields
(up to 10~T) and at elevated temperatures (up to room
temperature).

V.V.F. and I.K.Y. acknowledge the financial support by the
European Community grant INTAS-94-3562.

\end{document}